\documentstyle[aps,preprint]{revtex}

\begin{document}

\author{
Daniele Passerone$^{a,b}$\cite{corrauth}, Furio Ercolessi$^{a,b}$\cite{aff},
 and Erio Tosatti$^{a,b,c}$\cite{aff2} \\
{\it a) Istituto Nazionale per la Fisica della Materia (INFM)}\\
{\it b) International School for Advanced Studies, Trieste, Italy}\\
{\it c) The Abdus Salam International Center for Theoretical Physics,}\\
{\it Trieste, Italy}
}
\title
{Islands, craters, and a moving surface step on a hexagonally reconstructed
(100) noble metal surface}
\maketitle
\begin{abstract}
Deposition/removal of metal atoms on the hex reconstructed (100) 
surface of Au, Pt and Ir should present intriguing aspects, since a new island 
implies hex $\rightarrow$ square deconstruction of the substrate, and a new 
crater the square $\rightarrow$ hex reconstruction of the uncovered layer. To obtain a microscopic understanding of 
how islands/craters form in these conditions, we have conducted simulations of 
island and crater growth on Au(100), whose atomistic behavior, including 
the hex reconstruction on top of the square substrate, is well described by means of 
classical many-body forces. By increasing/decreasing the Au coverage on Au(100), 
we find that island/craters will not grow unless they exceed a critical size of about 8-10 
atoms. This value is close to that which explains the nonlinear coverage dependence observed 
in molecular adsorption on the closely related surface Pt $(100)$. This threshold size is
rationalized in terms of a transverse step correlation length,
 measuring the spatial extent where reconstruction of a given plane is disturbed by the 
nearby step. 
\end{abstract}
\section{Introduction}

The heavy noble metal $(100)$ surfaces possess the so-called
hex-reconstruction, where the top monolayer spontaneously converts from
square to (approximately) hexagonal (in fact, triangular), with 
a lateral density increase of $25-30\%$ \cite{xraydensity}. The 
second atomic layer, immediately below the first, remains instead 
square, with a bulk-like lateral density,
and only a minor local perturbation in correspondence of the
domain walls, or solitons, which the hex layer forms by (incommensurate)
epitaxy onto it.

Let us imagine the flat, hex-reconstructed $(100)$ surface and consider what should happen if we were to ideally deposit one further 
monolayer on top of it. The new top monolayer should be itself 
hex-reconstructed, because that is the lowest energy configuration
for this surface. However the former top layer, now covered and 
turned into a second layer, must {\it de-}construct, from the hex 
state back to a square lattice. Conversely, we
may imagine removing the top monolayer. The former, unreconstructed second
layer must now acquire some extra atoms, in order to become 
hex-reconstructed, again because that is the lowest energy state.

The questions now are: how exactly should all this happen? Where 
do the excess atoms, expelled from the covered layer, go? Conversely, 
where do the extra atoms needed for top layer reconstruction come from? 
And what other consequences does this peculiar situation have? 
A related problem which provided some inspiration for this work is that of 
the surprising nonlinearity observed, mainly by 
King {\it et al. }\cite{king}
in surface adsorption of molecules ({\rm CO, O}$_2,$ {\rm D}$_2$) versus
coverage,  on a hex reconstructed {\rm Pt }$(100)$ substrate. 

In this paper we describe
work which lays the ground for addressing some of these questions. 

We carried out Molecular Dynamics (MD) simulation work addressing
specifically the hex-reconstructed {\rm Au }$(100).$ 
  The temperature dependence
  of the top layer density allows us to study
  phenomena associated with density changes by simply changing the
  sample temperature in a particle-conserving system:
upon increasing temperature lateral density of a flat reconstructed 
Au $(100)$ surface shows a tendency to increase. We find in our simulations \cite{dati} an increase from the $T=0$ 
lateral density of $1.24$ with respect to the bulk, to $1.35$ for $T=1100 K$. This behavior is 
in close agreement with experiment for both Au and Pt \cite{xraydensity}.
Thus,
heating is equivalent to removing atoms, cooling to adding atoms.

The work done so far includes the following:

\begin{itemize}
\item  interplay of step and hex reconstruction, showing their important
mutual influence;

\item  sudden formation of a small adsorbed island (spontaneously expelled
by heating) with accompanying deconstruction of the covered portion;

\item  sudden formation of a small crater (spontaneously formed by cooling)
with accompanying reconstruction of the uncovered substrate portion;

\item  sudden step retraction (obtained by heating) with reconstruction of
the uncovered substrate portion, via ``incorporation'' of step edge atoms.
\end{itemize}

In the following, we shall briefly summarize some of our results, leaving a more
proper and detailed account for a separate publication.

\section{Method}

The hexagonal reconstruction of Au, Pt and Ir(100) consists of a
spontaneously stabilized 2D close-packed monolayer on top of the
otherwise (100) crystal. This phenomenon is reasonably well understood, but quite hard
to handle at the electronic level \cite{takeuchi89}. We found quite
some time ago that it was possible to reproduce it with quantitative
accuracy for Au(100)\cite{furio86}, using 
classical potentials of the many-body type, the glue\cite{furio88}
potentials, 
very specially and carefully optimized for Au. Within that approximation, it is possible 
to carry out extensive
simulations of some of the the situations imagined above, and obtain
from them a microscopic insight into those otherwise puzzling
questions.

With the glue potential for gold, we carried out classical MD simulation.
Newton's equations were integrated numerically,
allowing for large length-scales (at least 50 \AA\ lateral size) 
and long simulation times, (at least 1 nsec),
working at sufficiently high temperature so as to attain sufficient
atom mobilities. 

The geometry chosen for simulating our surfaces is an N-layer (001)
slab, (N=12-16) with periodic boundary conditions (PBC) along the (100)
and (010) directions. The top (001) surface was free, whereas the bottom
one consisted of 3 frozen bulk-like layers. The typical total atom number ranged from 20000 to 30000 \footnote {The free $(001)$ surface of Au is generally reconstructed,
with a periodicity which depends on temperature, and is generally incommensurate \cite{watson}. Our cell can only accommodate a commensurate periodicity (at least in the absence of steps) and we choose that to be $(5 \times 1)$, or $(5 \times 25)$, rather close
 to the actual one, $(5 \times 34)$. We do not expect these small deviations to 
be very important. An additional aspect is that of rotations of the reconstructed 
overlayer. The experimental rotational angle is given as $0.84^\circ$ \cite{xraydensity}, jumping to zero at $1000 K$. Rotations are basically incompatible with 
periodic boundary conditions, and have been neglected.}. 

In order to simulate a surface with a {\em single} step, we generated
new suitable PBCs which transform an $A$ layer into a $B$ layer 
(within an ABAB.. (100) stacking sequence) when crossing
the slab boundary in the direction orthogonal to the step.
This artifact allows the study of an isolated step which interacts only
with its repeated image, and the terrace size remains constant during
the system evolution.

Temperature was controlled, and the system was carefully and gradually heated 
through velocity rescaling. Mobility of surface atoms became
non-negligible for $T>1000K$ (well below melting, $T_m =1336 K$ for Au). 

The simulation of atom addition and removal represents a difficult task
in the canonical ensemble. Working fully canonically, i.e., conserving
particles, we were able to obtain a similar outcome by exploiting 
a peculiar feature of the $(100)$
hex reconstructed surface, namely the fact that its lateral density increases 
with temperature, notably by about 5\% from 700 to 1000 K. \cite{dati}

When a step is present, as is the case here, it will retract upon heating to accommodate for this
top layer density increase. The net movement of the step in fact 
provides in this way for us a very natural method to gauge the optimal spontaneous change 
of surface lateral density. Moreover, we can observe how the lateral
density varies locally, depending on the planar coordinate relative to the
step position.

If we consider now a flat, step-free surface, and we heat it up 
suddenly, the associated increase
of optimal lateral hex layer density will induce a strong tensile 
surface stress, which cannot be relieved in the absence of a 
defect. If that stress overcomes a certain critical limit,  
we can expect the sudden formation of one or more craters, leading to 
a state quite similar to that which we could have obtained by removing 
atoms at fixed temperature. Conversely, sudden cooling should lead
to compressive stress, and eventually islands of excess atoms will pop up
to relieve that stress.

This method is completely ad hoc for our situation, and it can only
work in practice if the hex layer is relatively free to slide parallel
to itself. Luckily, we found that this is the case for gold in the
temperature range 800-1000K used here.

An alternative, and certainly more standard way to add or remove atoms
in order to study island or crater formation would be Grand Canonical 
Monte Carlo (GCMC). Although we did succeed in implementing it
for certain Au surfaces, we eventually found the MD technique
described above more useful for the present purposes. GCMC is 
in the first place very difficult
to equilibrate, and moreover it does not provide as much desirable information
on the dynamics, as MD does.

Conversely, GCMC can be of great help in all cases where a 
strong density change (like in the square$\rightarrow$hex transformation)
must be handled. We will demonstrate how 
that works for Au $(100)$ in a forthcoming paper \cite{GCMC}.

\section{A step on the hex Au $(100)$ surface}

We simulated a $(100,1,1)$ vicinal surface of Au $(100)$, with a single step and
a wide terrace (maximum terrace size was $50 \times 50$ nearest neighbor
distances). Fig. 1 (upper part) shows the side view of the surface with a step.

The atoms of the second layer are marked in white; we note that the part 
of the second layer covered by the terrace is unreconstructed (A) whereas
the (B) part is reconstructed. If we follow the lateral 
density of the second layer from A to B, we will cross a transition zone 
in correspondence with the step. The width of this zone marks the surface correlation
length as probed by 
the step, and can be
extracted from MD simulations, and the result for the temperature of $T=900 K$ are
shown in Fig. 1 (lower part). The correlation
length is roughly  $5$ \AA\ at $900 K$ and increases to $10$ \AA\ at $1250 K$ (not shown). 
This result gives a measure of the influence of the step on the lateral 
coordination in its neighborhood, and reveals an interesting interplay 
between step and reconstruction.

\section{Island formation}

The reconstructed $(100)$ surface of noble metals undergoes an order $\rightarrow$disorder 
transition at about $T=.8\,T_m$. Low temperature deconstruction can, however, be induced upon adsorption of 
molecular species. The Cambridge group \cite{king} has carried out thorough studies of adsorption of light 
molecules such as CO, D$_2$ and O$_2$ onto a fully reconstructed Pt $(100)$ surface. They found 
that molecular island form, but that the
growth rate of the islands increases extremely slowly at 
low concentrations, roughly like the 
fourth power of molecular coverage (we shall call this King's law).
The explanation offered for this phenomenon is that while
the hex-reconstructed surface is unreactive, and will bind the 
molecules, the opposite is true for the {\it deconstructed}
,
square $(100)$ surface. The latter however must nucleate (under the island), and this requires a finite island size, of no less
than 4 ad-molecules.

It should be noted that King's exponent of
about $4$ implies that about $8$
Pt atoms must switch from hex to square for the adsorption process to grow. Hence
King's law is most likely telling us a property of the clean surface.
This is confirmed by the circumstance that the exponent is not very dependent on 
the adsorbed species.
The next observation is that very much the same deconstruction
must take place with homoepitaxy. Hence we expect that upon deposition of
Pt on Pt$(100)$, or on Au$(100)$, particularly at high temperatures when
equilibrium can be established, there should be a minimum critical island
size, of order 8 or so atoms, related to deconstruction of the
substrate.

We mimicked homogeneous atoms addition/removal through the 
already mentioned temperature jump technique. An alternative method was to prepare the 
system with a certain surface excess density, 
and to wait for the excess atoms to form an island.
In both cases, the island/crater growth was mainly determined 
by the density difference between square and hexagonal phase and
not by the initial conditions of the simulation; the 
dynamics can be therefore trusted as true island/crater growth dynamics,
whatever the method used to determine lateral density excess/deficit.

In this section we will focus on island formation case, adding atoms at a temperature of
$1200 K$ in order to have sufficient mobility of surface atoms.  

An initial excess density of 0.06 $\rho_b$ ($\rho_b$ being the lateral density of a bulk $(100)$ layer) at the surface causes the 
appearance of small fluctuating islands. However these islands are readsorbed quickly
by the substrate so long as their size is smaller than about 10 atoms. If the
size is greater than 10-15 atoms, however, 
the island is not 
readsorbed and begins to grow.
Figure 2(a) shows the time evolution of the maximum island size on the surface; there is clearly a critical size above which the island size grows.

Our explanation for this critical size is the following:
as long as the island is small, there is no deconstruction under it.
In other words if the border of the island has the same role of the
single step 
described in the previous section, the island size is too small.
 No deconstruction occurs until
the diameter of the island is greater than the 
surface correlation length as felt by the step.
When the size of the island exceeds this value, only then 
the substrate can deconstruct and growth can continue, now fueled by the excess atoms ejected by the lower layer. These atoms correspond 
to the density difference between
hexagonal and square order under the island, and they make
the island growth very fast. The deconstruction is 
almost completed when the island has a size of about 25-30 atoms, as 
shown in a snapshot of the simulation (t=175 ps), shown in Figure 3.

\section{Crater formation}

A symmetric situation with respect to the last section is the formation
of craters on a flat surface. 
We used temperature as the driving force for the density 
change at the surface, by increasing temperature from 800 to $T=950K$. At this 
temperature, surface atom mobility is sufficiently high. 
The main results concerning craters are the following:

\begin{itemize}
\item  The formation of craters requires a slightly smaller nucleus with a
critical size $N_H$ of about $8-10$ atoms.
Figure 2(b) shows this behaviour; the size (in atoms) of a crater
is plotted versus the simulation time. At a size of about 8, a jump occurs and the
 crater growth subsequently continues linearly, up to 
saturation.
This jump is associated with reconstruction of the crater bottom.


\item  The mechanism for further growth of the craters with $N>N_H$ is
the following: atoms in the hole, initially arranged in a square lattice,
are undercoordinated; atoms at the boundary of the crater are ``eaten up''
increasing the dimension of the hole. The growth is less dramatic than that 
of the island.

\end{itemize}

Summarizing, there appears to be a connection between the critical nucleus for 
growth of craters and of islands and the reconstruction correlation length as probed
by a step on a surface.
We can infer from the two situations we have examined (the crater and
the island) that reconstruction and deconstruction play a crucial role in determining their onset.
The size of the critical nucleus predicted for Au $(100)$ is of 8-10 atoms, in 
remarkably close agreement with the 8-atom size which can be extracted by King's exponent of
4 for molecular adsorption on Pt $(100)$

\section{Discussion}
We have found that reconstruction/deconstruction introduces a natural
critical size for the nucleation of islands and craters of Au on Au (100).
This size does not show the normal dependence upon supersaturation, expected of normal
nucleation processes, and appears more as an intrinsic characteristics of that 
surface. The reason can be as follows. The nucleation free energy barrier as 
a function of increasing size has a nonstandard shape,
with a large, sudden drop around $\simeq 15$ atoms, when substrate reconstruction/deconstruction can occur. That drop has the effect of {\it pinning} the critical 
size, making it independent of sovrasaturation and, possibly, also of the adsorbed 
species.
Figure 4 shows a very schematic representation of this point in the crater case. For small craters, the bottom of the crater is not reconstructed. At the radius of about 6 Angstrom, deconstruction can occur and a jump in the free energy is observed; by changing the sovrasaturation, the jump position (which depends only on the interplay between step and reconstruction), does not change.

\section*{Acknowledgments}
We thank D. A. King and Bo Persson for fruitful and constructive discussions. 
Work of D. P. at SISSA is directly supported by MURST.

\section*{Figure captions}
\begin{itemize}
\item Figure 1. Upper part: side view of a surface with a single step. The second layer atoms are drawn in white. (A) denotes the part of the second layer which lies under the terrace and is unreconstructed, (B) denotes the uncovered 
reconstructed terrace. Lower part: profile of the lateral density in the ``second layer'' 
for $T=900 K$. The layer has a square structure to the left, where it is covered by a terrace, and a dense hexagonal structure to the right. 
The interface between the two zones is smoothened by the presence
of a finite correlation length of the step. At $T=1200 K$ (not shown) the profile
is smoother, and the correlation length larger.
\item Figure 2. (a): the maximum island size for an excess density of about $0.06$. There is a jump in the island size and a growth up to a saturation value of
 about 38 atoms, corresponding to the initial excess density. The arrow indicates the onset of deconstruction of the substrate under the island. This deconstruction is completed at a size of about 20.
(b): the crater formation case; time evolution of a crater size. Please note the jump at
a critical value for the size.
\item Figure 3. (color) Snapshot of the island growth simulation after 175 $ps$. Red atoms are adatoms. The view from the bottom shows the almost complete 
deconstruction under the red island, bigger than the critical size.
\item Figure 4. Schematic profile of Gibbs free energy (at different chemical potential) upon formation of a crater of radius $r$. The jump occurs at the crater reconstruction, and does not depend on sovrasaturation.

\end {itemize}

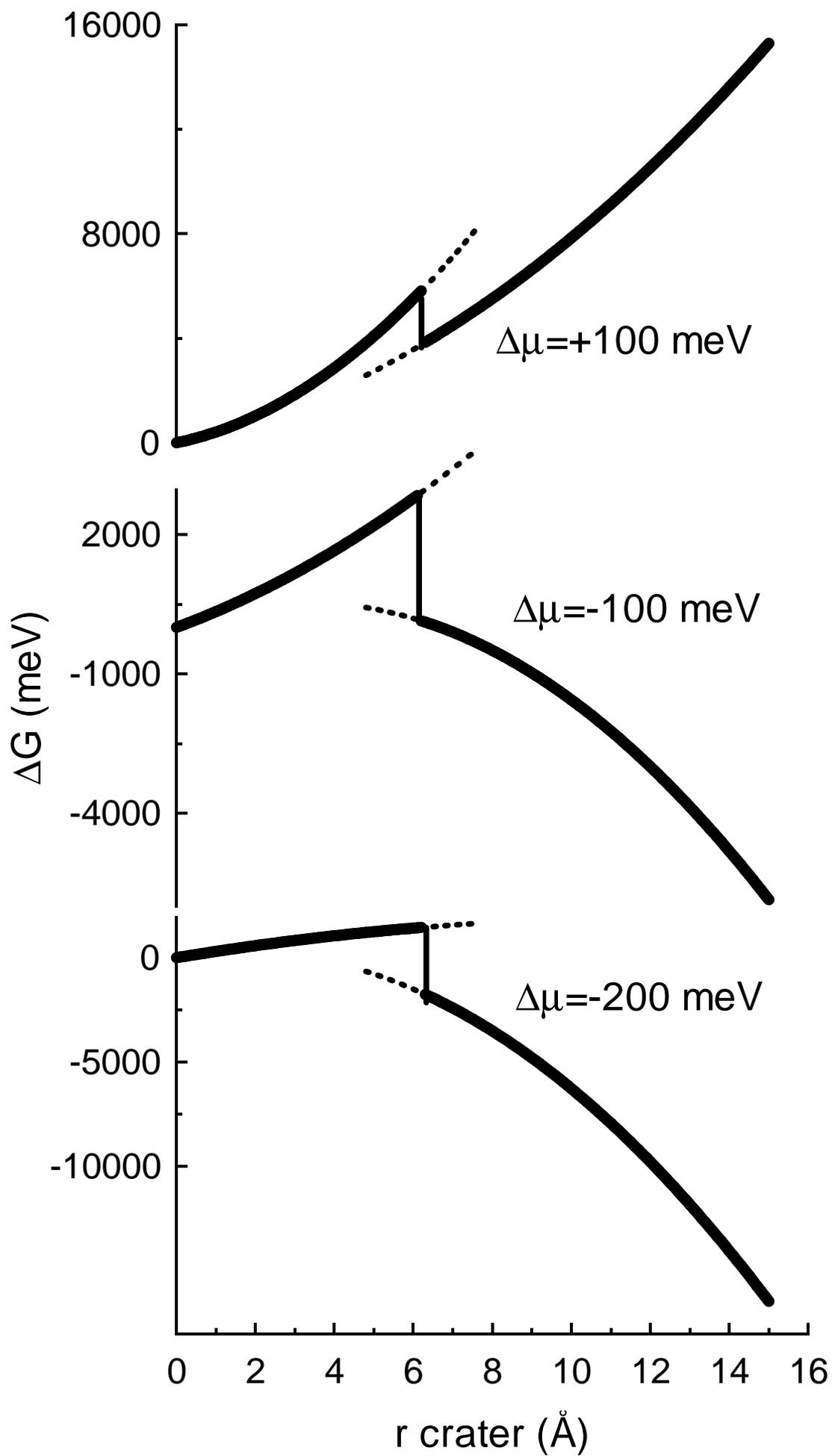
\end{document}